\documentstyle[aps,12pt,preprint]{revtex}
\draft

\title{Relativistic Charged Balls \thanks{%
The project supported by the Climbing-Up Program from Chinese Commission of Science and Technology. *liusiming@263.net}}
\author{YU Yun-qiang and LIU Si-ming$*$ \\
{\small Department of Physics, Peking University, Beijing 100871,China}}
\date{\today}
\begin{document}
\maketitle

\begin{abstract}
It is proven that the relativistic charged ball with its charge less than its mass (in natural units) cannot have a non-singular static configuration while its radius approaches  its external horizon size.  
This conclusion does not depend on the details of  charge distribution and the equation of state. The involved assumptions are (1) the ball is made of perfect fluid, (2) the energy density is everywhere non-negative.
\end{abstract}

\section{Introduction}
\label{usec1}

In 1959, Buchdahl$^{\cite{Buch}}$ first pointed out a relativistic effect
that a perfect fluid (uncharged) ball cannot have a non-singular static
configuration while its radius $R$ is not larger than 9/8 of its horizon size $R_{g}$ which is twice as large as its mass $M$ in natural units, if the energy
density is assumed not to increase outwards. For an incompressible
ball, pressure singularity will emerge at its center if $R$=(9/8)$R_{g}$.
In 1993, de Felice and Yu$^{\cite{Felice2}}$ found that if a configuration
with inner (central) singular boundary is acceptable, the shell shaped static
configuration can exist for any radius which is larger than $R_g$.
Although some theoretically favorable arguments have been discussed$^{\cite
{Felice3}\cite{Felice4}}$, the singular configuration seems exotic.

For charged balls, the pressure gradient will be partially balanced by the
electric force inside. Thus the central pressure of a dense configuration
will be weakened and the pressure singularity might be
avoided. In 1995, de Felice, Yu and Fang$^{\cite{Felice}}$ did find a series of
non-singular static configurations with $R$ arbitrarily approaching the corresponding horizon size $R_{+}=M+\sqrt{M^2-Q_0^2}$ in the case that $Q_0\to M$, where $Q_0$ is the charge of the ball.

The motivation of this work was to find the critical value of ratio $Q_{0}/M$
for the existence of series of regular static configurations with $R\to R_{+}$. Finally we find out that in case $Q_{0}/M$ is less than 1, no such
configuration exists. That is to say, $Q_{0}=M$ is the only case in which
the series of regular static configurations can contract to approach its horizon size.

In section \ref{usec2}, the general analysis of the problem is given. In section \ref{usec3}, we pose the theorem and prove it. In section \ref{usec4}, the physical meaning of the result is discussed.

\section{The Argument}
\label{usec2}

In the spherically symmetric case, the metric takes the form 
\begin{equation}
ds^2=-e^\nu dt^2+e^\lambda dr^2+r^2d\theta ^2+r^2\sin^2\theta d\varphi ^2. \label{u1}
\end{equation}
Here and here after the natural units system with $G=c=1$ is used. Assuming the
source is made of charged perfect fluid, Einstein's equations
for a static configuration will have the form$^{\cite{Beken}}$: 
\begin{eqnarray}
e^{-\lambda }\left(\frac{\lambda ^{\prime }}r-\frac 1{r^2}\right)+\frac 1{r^2} &=&\frac{%
Q^2}{r^4}+8\pi \rho   \label{uEin1}, \\
-e^{-\lambda }\left(\frac{\nu ^{\prime }}r+\frac 1{r^2}\right)+\frac 1{r^2} &=&\frac{Q^2%
}{r^4}-8\pi p  \label{uEin2}, \\
-\frac{e^{-\lambda }}2\left(\nu ^{\prime \prime }+\frac{\nu ^{\prime }{}^2}2+
\frac{\nu ^{\prime }-\lambda ^{\prime }}r-\frac{\nu ^{\prime }\lambda
^{\prime }}2\right) &=&-\frac{Q^2}{r^4}-8\pi p,  \label{uEin3}
\end{eqnarray}
where $\rho(r)$ and $p(r)$ are its energy density and isotropic
pressure respectively, and $Q(r)$ is the electric charge within the
radius $r$. The primes here stand for the derivatives with respect to $r$. In
general, an equation of state should be taken as input. Instead, we will
consider the energy density $\rho(r)$ as an arbitrary input. The charge distribution should depend on the
electromagnetic property of the medium. Since that property can be
arbitrarily assigned, we will consider $Q(r)$ as another arbitrary input.
Thus the set of equations (\ref{uEin1}) to (\ref{uEin3}) is a complete set for solving $\nu(r)$, $\lambda(r)$ and $p(r)$.

For convenience, we define a new variable $m(r)$ as
\begin{eqnarray}
m(r) &=&\int_0^r4\pi x^2\rho (x)dx+\frac 12\int_0^r\frac{Q^2(x)}{x^2
}dx+\frac{Q^2(r)}{2r} \label{u5}.
\end{eqnarray}
Let $r=R$ be the surface of the ball, the global parameters are
\begin{eqnarray}
M &=&m(R), \label{u6} \\
Q_0 &=&Q(R), \label{u7}
\end{eqnarray}
where $Q_0$ is the total charge of the ball. By using the variable $m(r)$, equation (\ref{uEin1}) can be solved out as 
\begin{equation}
e^{-\lambda(r)}=1-\frac{2m}r+\frac{Q^2}{r^2}.\label{u8}
\end{equation}
At the surface, it suits the Reissner-Nordstr\"om metric in standard form. Thus $M$ is called the total mass of the ball.

By eliminating pressure $p$ from equation (\ref{uEin2}) and (\ref{uEin3}), we have 
\begin{equation}
\left[e^{\nu -\lambda\over 2}\frac{\nu ^{\prime }}r\right]^{\prime }=\frac{1}{r^2}e^{
\lambda +\nu\over 2}\left(\frac{8Q^2}{r^3}-\frac{6m}{r^2}+8\pi \rho r\right).
\label{u9}
\end{equation}
As mentioned, $\rho(r)$ and $Q(r)$ are considered as inputs, then (\ref{u9}) is a second order differential equation for $\nu(r)$ only. We want our interior solution suits the exterior Reissner-Nordstr\"om metric at the surface. So the boundary conditions for $\nu(r)$ are as what follow:
\begin{eqnarray}
e^{\nu(R)}&=&1-\frac{2M}{R}+\frac{Q_0^2}{R^2}, \label{u10} \\
\left[e^{\nu(R)}\right]^{\prime}&\equiv&{\left[e^{\nu(r)}\right]^{\prime}}_{r=R}={2\over R}\left({M\over R}-{Q_0^2\over R^2}\right). \label{u11}
\end{eqnarray}

We will study the cases with $Q_0<M$ only.  In these cases, the exterior horizon size $R_+$ can be expressed by $Q_0$ and $M$ as
\begin{equation}
R_{+}=M+\sqrt{M^2-Q_0^2}. \label{u12}
\end{equation}
For assigned $R$, $Q_0$ and $M$, the solutions of equation (\ref{u9}) which satisfy the boundary conditions (\ref{u10}) and (\ref{u11}) will be called a Mathematical Solution Set (MSS).  Each element in the set
is defined by specified inputs of $\rho(r)$ and $Q(r)$. Surely, the solution is not always physically acceptable.  We assign the physically acceptable conditions (PAC) as what follow:
\begin{eqnarray}
R&>&R_+, \label{u13} \\
\rho(r)&\ge& 0, \label{u14} \\
e^{\lambda(r)}&>&0, \label{u15} \\
e^{\nu(r)}&>&0. \label{u16}
\end{eqnarray}
These are all necessary conditions. However, we will prove that while the radius $R$ approaches its corresponding horizon size $R_+$, there will be no element in MSS which satisfies PAC. By other words, for $Q_0<M$, there is a $R_0>R_+$ such that NO STATIC PHYSICAL CONFIGURATION WITH $R<R_0$  EXISTS.

     Since only the cases with $Q_0<M$ will be considered and only $R>R_+$  needs to be studied, we further define
\begin{eqnarray}
Q_0&\equiv&\sqrt{1-\Delta^2}M, \label{u17} \\
R&\equiv&(1+\epsilon)R_+\ =\ (1+\epsilon)(1+\Delta)M, \label{u18}
\end{eqnarray}
where $0<\Delta\le 1$ and $\epsilon>0$. Then the boundary values of $e^\nu$ and $e^\lambda$  are
\begin{eqnarray}
e^{\nu(R)} &=&\frac{\epsilon +(2+\epsilon )\Delta }{(1+\epsilon )^2(1+\Delta )}\epsilon, \label{u19} \\
e^{\lambda(R)}&=&e^{-\nu(R)},  \label{u20} \\ 
\left[e^{\nu(R)}\right]'&=&\frac{2(\epsilon +\Delta)}{(1+\epsilon)^3(1+\Delta)^2M} \label{u21}.
\end{eqnarray}
While $\Delta$ is fixed as a finite quantity and $\epsilon$ approaches zero, we see that $e^{\nu(R)}$ is infinitesimal, $[e^{\nu(R)}]'$ is finite and $e^{\lambda(R)}$ is infinite.

\section{The Proof}
\label{usec3}

The main theorem that we want to prove is as what follows:

THEOREM:  For $0<\Delta\le 1$, an $\epsilon_0$ (corresponding to $R_0$) can be found so that while $\epsilon<\epsilon_0$, any Element in MSS which satisfies PAC (\ref{u13}), (\ref{u14}) and (\ref{u15}) will always violates PAC (\ref{u16}).

      Some preparations are needed.

      For any element in MSS, integrating both sides of (\ref{u9}) from $r$ to $R$, we get
\begin{equation}
\left[e^{\frac \nu 2}\right]^{\prime }=\frac r2e^{\frac \lambda 2}\left[\frac{2}{R^2}\left({M\over R}-
\frac{Q_0^2}{R^2}\right)+F_{(r)}\right],  \label{u22}
\end{equation}
where the boundary conditions have been used and $F(r)$ is defined as
\begin{equation}
F(r)=-\int_r^R\frac{1}{x^2}e^{\frac{\lambda +\nu }2}\left[\frac{8Q^2}{x^3}-
\frac{6m}{x^2}+8\pi \rho x\right]dx.  \label{u23}
\end{equation}
We study the behavior of $e^{\nu(r)\over 2}$ and $F(r)$ from the boundary towards its center. At the boundary, we have $e^{\nu(R)}>0$ and $F(R)=0$. Therefore, there is an interval near $R$ in which
\begin{eqnarray}
e^{\nu(r)\over 2}&>&0, \label{u24} \\
F(r)&>&-{1\over R^2}\left({M\over R}-{Q_0^2\over R^2}\right). \label{u25}
\end{eqnarray}

LEMMA:  For any element of MSS which satisfies PAC (\ref{u13}), (\ref{u14}) and (\ref{u15}), an $\epsilon_1>0$ can be found such that while $\epsilon<\epsilon_1$, if (\ref{u24}) remains valid in $[\beta R, R]$ and (\ref{u25}) is valid in $(\beta R, R]$, it is impossible to have
\begin{equation}
F(\beta R)=-{1\over R^2}\left({M\over R}-{Q_0^2\over R^2}\right), \label{u26}
\end{equation}
where $\beta<1$ is a positive number.

We prove it in the following way: Suppose that $F(r)$ reaches $-{1\over R^2}\left({M\over R}-{Q_0^2\over R^2}\right)$ at $r=\beta R$, then an $\epsilon_1>0$ can be found such that while $\epsilon<\epsilon_1$, $e^{\nu(\beta R)\over 2}$ will be negative which contradicts (\ref{u24}).

PROOF OF THE LEMMA: In view of (\ref{u25}), (\ref{u22}) shows that $[e^{\nu(r)\over 2}]'>0$ for $r\in(\beta R,R]$. Then we have
\begin{equation}
0<e^{\nu(r)\over 2}\le e^{\nu(R)\over 2}\quad{\rm for\ }r\in[\beta R,R]. \label{u301}
\end{equation}
 
 Rewrite (\ref{u5}) as
\begin{equation}
(2rm-Q^2)'=4\pi\rho r^2-{Q^2\over 2r^2} \label{u27},
\end{equation}
then equation (\ref{u8}) leads to
\begin{equation}
\left[e^{-\frac{\lambda }2}\right]^{\prime }=e^{\frac \lambda 2}\left[\frac
m{r^2}-4\pi \rho r-\frac{Q^2}{r^3}\right]. \label{u28}
\end{equation}
On the other hand, the equations (\ref{u8}) and (\ref{u14}) show
\begin{eqnarray}
e^\lambda\ge 1, \label{u29} \\
1>{2m\over r}-{Q^2\over r^2}\ge 0. \label{u30}
\end{eqnarray}

By using formula (\ref{u28}), $F(r)$ turns to be
\begin{eqnarray}
F(r) &=&\int_r^R\frac{10}{x^2}e^{\frac \nu 2}d(e^{-\frac \lambda
2})+\int_r^R\frac{1}{x^2}e^{\frac{\nu +\lambda }2}\left[32\pi \rho x-\frac{4}{x^2}\left(m-
\frac{Q^2}{2x}\right)\right]dx.  \label{u31} 
\end{eqnarray} 
Integrating the first term of RHS by parts and using the boundary condition (\ref{u20}), we get
\begin{eqnarray}
F(r)&=&\frac{10}{R^2}e^{\nu(R)} -\frac{10}{r^2}e^{\frac{\nu(r)-\lambda(r)}2}
+20\int_r^R\frac{1}{x^3}e^{\frac{\nu -\lambda }2}dx  \nonumber \\
& &-10\int_r^R\frac{1}{x^2}e^{\frac{-\lambda}2}d(e^{\frac \nu 2})+\int_r^R
\frac{1}{x^2}e^{\frac{\nu +\lambda }2}\left[32\pi \rho x-\frac{4
}{x^2}\left(m-\frac{Q^2}{2x}\right)\right]dx.  \label{u32}
\end{eqnarray}
By taking away the positive terms in RHS, the equality turns to be an inequality:
\begin{equation}
F(r)>-\frac{10}{r^2}e^{\frac{\nu(r)-\lambda(r)}2}
-10\int_r^R\frac{1}{x^2}e^{\frac{-\lambda }2}d(e^{\frac \nu 2})
-\int_r^R\frac{4}{x^4}e^{\frac{\nu +\lambda }2}\left(m-\frac{Q^2}{2x}\right)dx \quad {\rm for}\ r\in[\beta R,R].\label{u33}
\end{equation}
Due to inequalities (\ref{u301}), (\ref{u29}), (\ref{u30}) and $r\ge\beta R$, we further have
\begin{equation}
F(r)>-{20\over (\beta R)^2}e^{\nu(R)\over 2}-{2\over (\beta R)^3}e^{\nu(R)\over 2}\int_r^R  e^{\lambda\over 2}dx \quad {\rm for}\ r\in[\beta R,R]. \label{u34}
\end{equation}

Suppose $F(r)$ reaches $-{1\over R^2}\left({M\over R}-{Q_0^2\over R^2}\right)$ at $r=\beta R$. Using (\ref{u34}), we get
\begin{eqnarray}
{1\over R^2}({M\over R}-{Q_0^2\over R^2})=-F(\beta R)<{20\over (\beta R)^2}e^{\nu(R)\over 2}+{2\over (\beta R)^3}e^{\nu(R)\over 2}\int_{\beta R}^R e^{\lambda\over 2}dr. \label{u35}
\end{eqnarray}

Now we consider the upper limit of $e^{\nu(\beta R)\over 2}$. By integrating (\ref{u22}) from $\beta R$ to $R$ and using the condition (\ref{u25}), we have
\begin{eqnarray}
e^{\nu(\beta R)\over 2}&=&e^{\nu(R)\over 2}
-\int_{\beta R}^R\frac r2e^{\frac\lambda 2}\left\{\frac{1}{R^2}\left({M\over R}-\frac{Q_0^2}{R^2}\right)
+\left[\frac{1}{R^2}\left({M\over R}-\frac{Q_0^2}{R^2}\right)+F(r)\right]\right\}dr \nonumber \\
&<&e^{\nu(R)\over 2}-{\beta\over 2R}\left({M\over R}-\frac{Q_0^2}{R^2}\right)\int_{\beta R}^R e^{\lambda\over 2}dr, \label{u36}
\end{eqnarray}
where $r\ge\beta R$ is also used. Solving out the integral $\int_{\beta R}^R e^{\lambda\over 2}dr$ from (\ref{u35}) and substituting the result into (\ref{u36}), we find inequality (\ref{u36}) becomes
\begin{eqnarray}
e^{\nu(\beta R)\over 2}
&<&e^{\nu(R)\over 2}-{\beta\over 2}\left({M\over R}-\frac{Q_0^2}{R^2}\right)\left[{{\beta}^3\over 2}e^{-\nu(R)\over 2}\left({M\over R}-\frac{Q_0^2}{R^2}\right)-10\beta\right]. \label{u37}  
\end{eqnarray}

So far, the radius $R$ is arbitrary. What we want to prove is that a $\epsilon_1$ (corresponding to $R_1$) does exist such that while $\epsilon<\epsilon_1$, the RHS of (\ref{u37}) will be always negative. This result implies the invalidity of (\ref{u26}). 

The negativity of the RHS of (\ref{u37}) means
\begin{equation}
4e^{\nu(R)}\left({M\over R}-{Q_0^2\over R^2}\right)^{-1}+20\beta^2e^{\nu(R)\over 2}
<\beta^4\left({M\over R}-{Q_0^2\over R^2}\right). \label{u38}
\end{equation}
Since it is a dimensionless formula, we express it by parameters $\Delta$ and $\epsilon$, that is
\begin{equation}
20\beta^2\sqrt{\epsilon(\epsilon+2\Delta+\epsilon\Delta)(1+\Delta)}(1+\epsilon)(\Delta+\epsilon)+4\epsilon(\epsilon+2\Delta+\epsilon\Delta)(1+\Delta)(1+\epsilon)^2
<\beta^4(\epsilon+\Delta)^2. \label{u39}
\end{equation}
For $\epsilon=0$, the LHS is zero and RHS is a positive finite quantity, so the inequality (\ref{u39}) is valid. Therefore, a positive $\epsilon_1$ exists such that (\ref{u39}) is valid for any $\epsilon<\epsilon_1$. That ends our proof. 
  
  COROLLARY: For $\epsilon<\epsilon_1$, if (\ref{u24}) is valid in the interval $[\beta R, R]$, then (\ref{u25}) must be valid in the same interval. 

There is no need to prove the corollary. We turn to prove the main theorem.

      PROOF OF THE THEOREM:  For any element in MSS which satisfies PAC (\ref{u13}),(\ref{u14}) and (\ref{u15}), we want to prove that if PAC (\ref{u16}) is also assumed to be valid in some interval $[\beta R,R]$, for small enough $\epsilon$, a contrary result will emerge. 

      By the corollary, the validity of PAC (\ref{u16}) leads to the validity of (\ref{u25}). Then we still have the inequality (\ref{u36}). Here we use $e^\lambda\ge 1$ to evaluate the integral in (\ref{u36}) as
\begin{equation}
\int_{\beta R}^R e^{\lambda\over 2}dr\ge (1-\beta)R. \label{u41}
\end{equation}
Substituting it into (\ref{u36}), we see that if 
\begin{equation}
2e^{\nu(R)\over 2}<\beta(1-\beta)\left({M\over R}-{{Q_0}^2\over R^2}\right) \label{u42}
\end{equation}
is valid, $e^{\nu(\beta R)\over 2}$ will certainly be negative. It is contrary to the validity of PAC (\ref{u16}). 

We rewrite inequality (\ref{u42}) by using parameters $\Delta$  and $\epsilon$
\begin{equation}
2(1+\epsilon)\sqrt{\epsilon(\epsilon+2\Delta+\Delta\epsilon)}
<{\beta(1-\beta)(\epsilon+\Delta)\over \sqrt{1+\Delta}}. \label{u43}
\end{equation}
Evidently, it is valid for $\epsilon=0$. Then $\epsilon_2$ can be chosen as
\begin{equation}
\epsilon_2={\beta^2(1-\beta)^2\Delta^2\over 16(1+\Delta)(1+3\Delta)}. \label{u44}
\end{equation}
It is not difficult to see, (\ref{u43}) will always be valid for $\epsilon<\epsilon_2$. Finally, we have to choose $\epsilon_0$ as
\begin{equation}
\epsilon_0=\min\{\epsilon_1,\epsilon_2 \} \label{u45}
\end{equation}
to assure the validity of the lemma and its corollary. However, such a $\epsilon_0$  does exist. It ends the proof of our theorem.  

\section{Discussions}
\label{usec4}

A. Pressure singularity.

As we know, for an incompressible perfect fluid (uncharged) ball, the pressure 
diverges and $e^{\nu}$ reaches zero at the center while $R={9\over 8}R_g$. If $R<{9\over 8}R_g$, the pressure singularity emerges at some $r_0>0$. While $R$ is smaller, the corresponding $r_0$ will be larger. In fact, the same thing happens in the charged case.

Equation (\ref{uEin2}) can be rewritten as 
\begin{equation}
\nu ^{\prime }=2\frac{4\pi rp+\frac m{r^2}-\frac{Q^2}{r^3}}{1-\frac{2m}r+
\frac{Q^2}{r^2}}.  \label{u46}
\end{equation}
For each configuration which satisfies PAC (\ref{u13})---(\ref{u15}) and has null $e^{\nu \over 2}$ inside, 
suppose
\begin{equation}
\lim_{r\to {r_0}_+}e^{\nu(r)\over 2}=0 \label{u461} 
\end{equation}
and $e^{\nu(r)\over 2}$ is positive for $r\in (r_0,R]$.
If there is an upper limit for pressure $p$ in $[r_0,R]$, from equation (\ref{u46}), 
we see that $\nu(R)-\nu(r)$ will be finite while $r\to {r_0}_+$.  
So $e^{\nu(r)}$ approaches a positive quantity while $r\to {r_0}_+$ which contradicts (\ref{u461}).
It implies that pressure $p$ must diverge at some point in $[r_0,R]$, and equation (\ref{u46}) shows
that $\nu^{\prime }$ diverges at the same point. However, equation 
(\ref{u22}) shows that $[e^{\frac \nu 2}]^{\prime }={1\over 2}e^{\nu\over 2}\nu'$ is finite in $[r_0,R]$, then we have that $e^{\nu(r)\over 2}$ approaches zero while $p$ approaches infinity. Because $e^{\nu\over 2}$ is a positive function in $(r_0,R]$, we deduce that pressure approaches infinity at the same point where $e^{\nu \over 2}$ approaches zero for the first time from the boundary inwards.

Qualitatively, pressure increases inwards from $p(R)=0$ and the electric force tends
to weaken the increasing of pressure. As we expected, this effect may avoid the emergence of the pressure singularity. Our theorem shows that while $Q_0<M$, the electric force is not strong enough to avoid the pressure singularity. 

\medskip
\noindent B. The case of $Q_0=M$.

      In this case, our proof fails to be valid since the $\epsilon_0>0$ cannot be found. In fact, the opposite can be proved. If the given $\rho(r)$ and $Q(r)$ satisfy 
\begin{eqnarray}
e^{-\lambda }(r) &\ge &1-\frac{2r^2}{R^2}\left({M\over R}-{Q_0^2\over 2R^2}\right),
\label{u47} \\
\frac{6m}{r^2} &\le &\frac{8Q^2}{r^3}+8\pi r\rho,   \label{u48}
\end{eqnarray}
elements in MSS satisfying PAC (\ref{u13})---(\ref{u16}) do exist for $R\to R_+$. The solution found by de Felice et al. offers an example. Other series of models can also be produced. However, it is worthy of noting that PAC (\ref{u13})---(\ref{u16}) are necessary conditions only. So it does not imply that static physical configurations do exist while $R\to R_+$. 

\medskip
\noindent C. The involved assumptions.

We conclude that no static physical configuration exists for $R\to R_+$ if $Q_0<M$. Here only some naive assumptions are used. They are 

1: the charged ball is made of perfect fluid, 

2: The energy density $\rho(r)$ is non-negative everywhere.

\noindent No particular assumption for property of the medium is involved.

\end{document}